\documentclass[12-pt]{article}
\usepackage{amssymb,amsmath,mathptmx,makeidx,graphicx,url}
\DeclareMathOperator{\KS}{\textit{C}\hspace*{0.5pt}}
\DeclareMathOperator{\KP}{\textit{K}\hspace*{0.5pt}}
\DeclareMathOperator{\KT}{\textit{CT}\hspace*{0.5pt}}
\newtheorem{theorem}{Theorem}

\newcommand{\cnd}{\hspace*{1pt}|\hspace*{1pt}}

\begin{document}
\title{Algorithmic statistics revisited}
\author{Nikolay Vereshchagin\thanks{Moscow State University and Yandex, \texttt{nikolay.vereshchagin@gmail.com}}, Alexander Shen\thanks{LIRMM  CNRS \&UM (Montpellier), on leave from IITP RAS, \texttt{alexander.shen@lirmm.fr}}}
\date{}
\maketitle

\begin{abstract}
The mission of statistics is to provide adequate statistical hypotheses (models) for observed data. But what is an ``adequate'' model? To answer this question, one needs to use the notions of algorithmic information theory. It turns out that for every data string $x$ one can naturally define  ``stochasticity profile'', a curve that represents a trade-off between complexity of a model and its adequacy. This curve has four different equivalent definitions in terms of (1)~randomness deficiency, (2)~minimal description length, (3)~position in the lists of simple strings and (4)~Kolmogorov complexity with decompression time bounded by busy beaver function. We present a survey of the corresponding definitions and results relating them to each other.
\end{abstract}

\section{What is algorithmic statistics?}

The laws of celestial mechanics allow the astronomers to predict the observed motion of planets in the sky with very high precision. This was a great achievement of modern science---but could we expect to find equally precise models for all other observations? Probably not. Thousands of gamblers spent all theirs lives and their fortunes trying to discover the laws of the roulette (coin tossing, other games of chance) in the same sense---but failed. Modern science abandoned these attempts. It says modestly that all we can say about the coin tossing is the statistical hypothesis (model): \emph{all trials are independent and \textup(for a fair coin\textup) both head and tail have probability $1/2$}. The task of mathematical statistics therefore is to find an appropriate model for experimental data. But what is ``appropriate'' in this context?

To simplify the discussion, let us assume that experimental data are presented as a bit string (say, a sequence of zeros and ones corresponding to heads and tails in the coin tossing experiment). We also assume that a model is presented as a probability distribution on some finite set of binary strings. For example, a fair coin hypothesis for $N$ coin tossings is a set of all strings of length $N$ where all elements have the same probability $2^{-N}$.  Restricting ourselves to the simplest case when a hypothesis is some set $A$ of strings with uniform distribution on it, we repeat our question:
\begin{quotation}
Assume that a bit string $x$ (data) and a set $A$ containing $x$ (a model) are given; when do we consider $A$ as a good ``explanation'' for $x$?
\end{quotation}

Some examples show that this question cannot be answered in the framework of classical mathematical statistics. Consider a sequence $x$ of $100$ bits (the following example is derived from the random tables~\cite{random-tables}):
\begin{quote} 
    01111 10001  11110 10010  00001 00011  00001 10010  00010 11101\\
    10111 11110  10000 11100  00111 00000  01111 01100  11011 01011   
\end{quote}  
Probably you would agree that the statistical hypothesis of a fair coin (the set $A=\mathbb{B}^{100}$ of all $100$-bit sequences) looks as an adequate explanation for this sequence. On the other hand, you probably will not accept the set $A$ as a good explanation for the sequence $y$:
\begin{quote} %
    00000 00000  00000 00000  00000 00000  00000 00000  00000 00000\\
    00000 00000  00000 00000  00000 00000  00000 00000  00000 00000   
\end{quote} 
but will suggest a much better explanation $B=\{y\}$ (the coin that always gives heads). On the other hand, set $C=\{x\}$ does not look like a reasonable explanation for $x$. How can we justify this intuition?

One could say that $A$ is not an acceptable statistical hypothesis for $y$ since the probability of $y$ according to $A$ is negligible ($2^{-100}$). However, the probability of $x$ for this hypothesis is the same, so \emph{why is $A$ acceptable for $x$} then? And if $B$ looks like an acceptable explanation for $y$, \emph{why $C$ does not look as an acceptable explanation for $x$}? 

The classical statistics, where $x$ and $y$ are just two equiprobable elements of $A$, cannot answer these questions. Informally, the difference is that $x$ looks like a ``random'' element of $A$ while $y$ is ``very special''. To capture these difference, we need to use the basic notion of algorithmic information theory, Kolmogorov complexity,\footnote{We assume that the reader is familiar with basic notions of algorithmic information theory and use them freely. For a short introduction see~\cite{uppsala-notes}; more information can be found in~\cite{li-vitanyi}.} and say that $x$ has high complexity (cannot be described by a program that is much shorter than $x$ itself) while $y$ has small complexity (one can write a short program that prints a long sequence of zeros). This answers our first question and explains why $A$ could be a good model for $x$ but not for $y$. 

Another question we asked: why $B$ is an acceptable explanation for $y$ while $C$ is not an acceptable explanation for $x$? Here we need to look at the complexity of the model itself: $C$ has high complexity (because $x$ is complex) while $B$ is simple (because $y$ is simple). 

Now let us consider different approaches to measuring the ``quality'' of statistical models; they include several parameters and a trade-off between them arises. In this way for every data string $x$ we get a curve that reflects this trade-off. There are different ways to introduce this curve, but they are all equivalent with $O(\log n)$ precision for $n$-bit strings. The goal of this paper is to describe these approaches and equivalence results.

\section{$(\alpha,\beta)$-stochastic objects}
\label{sec:stochastic}

Let us start with the approach that most closely follows the scheme described above. Let $x$ be a string and let $A$ be a finite set of strings that contains $x$. The ``quality'' of $A$ as a model (explanation) for $x$ is measured by two parameters:
\begin{itemize}
\item the Kolmogorov complexity $\KS(A)$ of $A$;
\item the randomness deficiency $d(x\cnd A)$ of $x$ in $A$. 
\end{itemize}
The second parameter measures how ``non-typical'' is $x$ in $A$ (small values mean that $x$ looks like a typical element of $A$) and is defined as
    $$
d(x\cnd A)= \log\#A - \KS(x\cnd A).
    $$
Here $\log$ stands for binary logarithm, $\#A$ is the cardinality of $A$ and $\KS(u\cnd v)$ is the conditional complexity of $u$ given $v$. Using $A$ as the condition, we assume that $A$ is presented as a finite list of strings (say, in lexicographical ordering). The motivation for this definition: for all $x\in A$ we have $\KS(x\cnd A)\le \log\#A+O(1)$, since every $x\in A$ is determined by its ordinal number in $A$; for most $x\in A$ the complexity $\KS(x\cnd A)$ is close to $\log\#A$ since the number of strings whose complexity is much less than $\log\#A$, is negligible compared to $\#A$. So the deficiency is large for strings that are much simpler than most elements of $A$.\footnote{There is an alternative definition of $d(x\cnd A)$. Consider a function $t$ of two arguments $x$ and $A$, defined when $x\in A$, and having integer values. We say that $t$ is \emph{lower semicomputable} if there is an algorithm that (given $x$ and $A$) generates lower bounds for $t(x,A)$ that converge to the true value of $t(x,A)$ in the limit. We say that $t$ is a \emph{probability-bounded test} if for every $A$ and every positive integer $k$ the fraction of $x\in A$ such that $t(x,A)>k$ is at most $1/k$. Now $d(x\cnd A)$ can be defined as the logarithm of the maximal (up to $O(1)$-factor) lower semicomputable probability-bounded test.}

According to this approach, a good explanation $A$ for $x$ should make both parameters small: $A$ should be simple and $x$ should be typical in $A$. It may happen that these two goals cannot be achieved simultaneously, and a trade-off arises. Following Kolmogorov, we say that $x$ is \emph{$(\alpha,\beta)$-stochastic} if there exists $A$ containing $x$ such that $\KS(A)\le \alpha$ and $d(x\cnd A)\le \beta$. In this way we get an upward closed set 
$$
S (x) =\{ \langle \alpha,\beta\rangle\mid \text{$x$ is $(\alpha,\beta)$-stochastic}\}
$$
If $x$ is a string of length $n$, the set $A$ of all $n$-bit strings can be used as a description; it gives us 
the pair $(O(\log n), n-\KS(x)+O(\log n))$ in $S(x)$. Indeed, we can describe $A$ using $O(\log n)$ bits and the deficiency is $n-\KS(x\cnd A)=n-\KS(x\cnd n)=n-\KS(x)+O(\log n)$. On the other hand, 
there is a set $A\ni x$ of complexity $\KS(x)+O(1)$ and deficiency $O(1)$ (namely, $A=\{x\}$). 
So the boundary of the set $S(x)$ starts below the point $(0,n-\KS(x))$ and decreases to $(\KS(x),0)$ for arbitrary $n$-bit string $x$, if we consider $S(x)$ with $O(\log n)$ precision.\footnote{As it is usual in algorithmic information theory, we consider the complexities up to $O(\log n)$ precision if we deal with strings of length at most $n$. Two subsets $S,T\subset\mathbb{Z}^2$  are the same for us if $S$ is contained in the $O(\log n)$-neighborhood of $T$ and vice verse.}

The boundary line of $S(x)$ can be called a \emph{stochasticity profile} of $x$. As we will see, the same curve appears in several other situations.

\section{Minimum description length principle}
\label{sec:mdl}

Another way to measure the ``quality'' of a model starts from the following observation: if $x$ is an element of a finite set $A$, then $x$ can be described by providing two pieces of information:
\begin{itemize}
\item the description of $A$;
\item the ordinal number of $x$ in $A$ (with respect to some ordering fixed in advance).
\end{itemize}
This gives us the inequality
    $$
\KS(x)\le \KS(A)+\log \#A 
    $$
that is true with precision $O(\log n)$ for strings $x$ of length at most $n$.\footnote{The additional term $O(\log\KS(A))$ should appear in the right hand side, since we need to specify where the description of $A$ ends and the ordinal number of $x$ starts, so the length of the description ($\KS(A)$) should be specified in advance using some self-delimiting encoding. One may assume that $\KS(A)\le n$, otherwise the inequality is trivial, so this additional term is $O(\log n)$.}

The quality of the hypothesis $A$ is then measured by the difference
$$
\delta(x,A)=\KS(A)+\log\#A-\KS(x)
$$
 between the sides of this inequality. We may call it ``optimality deficiency'' of $A$, since it shows how much do we lose in the length of the description if we consider two-part description based on $A$ instead of the best possible one. For a given string $x$ we can then consider the set $O(x)$ of pairs $\langle\alpha,\beta\rangle$ such that $x$ has a model of complexity at most $\alpha$ and optimality deficiency at most $\beta$.

\begin{theorem}\label{th:stochastic-optimal}
For every string $x$ of length at most $n$ the sets $S(x)$ and $O(x)$ coincide with $O(\log n)$-precision: each of them is contained in the $O(\log n)$-neighborhood of the other one.
\end{theorem}

Speaking about neighborhoods, we assume some standard distance on $\mathbb{R}^2$ (the exact choice does not matter, since we measure the distance up to a constant factor).

Let us note now that in one direction the inclusion is straightforward. A simple computation shows that the  randomness deficiency is always less than 
the optimality deficiency \emph{of the same model} (and the difference between them equals $\KS(A\cnd x)$, where $A$ is this model).

The opposite direction  is more complicated: a model with small randomness deficiency may have large optimality deficiency.  This may happens when $\KS(A\cnd x)$ is large.\footnote{%
Let $x$ and $y$ be independent random strings of length $n$, so the pair $(x,y)$ has complexity close to $2n$. Assume that $x$ starts with $0$ and $y$ starts with $1$.  Let $A$ be the set of strings that start with $0$, plus the string $y$. Then $A$, considered as a model for $x$,  has large optimality deficiency but small randomness deficiency. To decrease the optimality deficiency, we may remove 
$y$ from $A$.} 
However, in this case we can find another model and decrease the optimality deficiency as needed:  \emph{for every string $x$ and every model $A$ for $x$} (a finite set $A$ that contains $x$) \emph{there exists another model $A'$ for $x$ such that $\log\#(A')=\log\#A$ and $\KS(A')\le \KS(A)-\KS(A\cnd x)+O(\log n)$}, where $n$ is the length of~$x$. This result looks surprising at first, but note that if $\KS(A\cnd x)$ is large, then there are many sets $A'$ that are models of the same quality (otherwise $A$ can be reconstructed from $x$ by exhaustive search). These sets can be used to find $A'$ with required properties.

\medskip
The definition of the set $O(x)$ goes back to Kolmogorov~\cite{Ko74a}; however, he used a slightly different definition: instead of $O(x)$ he considered 
the function 
$$
   h_{x}(\alpha) = \min_{A} \{\log \# A : x\in A,\; \KS(A) \leq \alpha\},
$$
now called \emph{Kolmogorov structure function}. Both $O(x)$ and $h_x$ are determined by the set of all pairs $(\KS(A),\log\#A)$ for finite sets $A$ containing $x$, though in a slightly different way (since the inequality $\delta(x,A)\le\beta$ in the definition of $O(x)$ combines $\KS(A)$ and $\log\#A$). One can show, however, that the following statement is true with $O(\log n)$-precision for each $n$-bit string $x$: \emph{the pair $(\alpha,\beta)$ is in $O(x)$ if and only if $h_x(\alpha)\le \beta+\KS(x)-\alpha$}. So the graph of $h_x$ is just the boundary of $O(x)$ in different coordinates.

\begin{figure}
\begin{center}
\includegraphics[scale=1]{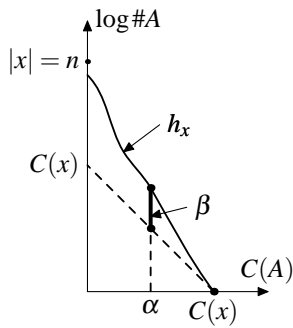}
\end{center}
\caption{The pair $(\alpha,\beta)$ lies on the boundary of $O(x)$ since the point $(\alpha,\KS(x)-\alpha+\beta)$ lies on the graph of $h_x$.}
\end{figure}

\section{Lists of simple strings}

We have seen two approaches that describe the same trade-off between the complexity of  a model and its quality: for every $x$ there is some curve (defined up to $O(\log n)$-precision) that shows how good can be a model with bounded complexity.  Both approaches gave the same curve with logarithmic precision; in this section we give one more description of the same curve.

Let $m$ be some integer. Consider the list of strings of complexity at most $m$. It can be generated by a simple algorithm: just try in parallel all programs of length at most $m$ and enumerate all their outputs (without repetitions). This algorithm is simple (of complexity $O(\log m)$) since we only need to know $m$.

There may be several simple algorithms that enumerate all strings of complexity at most $m$, and they can generate them in different orders. For example, two algorithms may start by listing all the strings of length $m-O(1)$ (they all have complexity at most $m$), but one does this in the alphabetical order and the other uses the reverse alphabetical order. So the string $00\ldots00$ is the first in one list and has number $2^{m-O(1)}$ in the other. 
But the distance  \emph{from the end of the list} is much more invariant:

\begin{theorem}\label{th:position-invariance}
Consider two programs of complexity $O(\log m)$ that both enumerate all strings of complexity at most $m$. Let $x$ be one of these strings. If there is at least $2^k$ strings after $x$ in the first list, then there is at least $2^{k-O(\log m)}$ strings after $x$ in the second list.
\end{theorem}

In this theorem we consider two algorithms that enumerate the same strings in different orderings. However, the Kolmogorov complexity function depends on the choice of the optimal decompressor (though at most by $O(1)$ additive term), so one could ask what happens if we enumerate the strings of bounded complexity for two different versions of the complexity function. A similar result (with similar proof) says that the change of an optimal decompressor used to define Kolmogorov complexity can be compensated by $O(\log m)$-change in the threshold $m$.

Now for every $m$ fix an algorithm of complexity at most $O(\log )m$ that enumerates all strings of complexity at most~$m$. Consider a binary string $x$; it appears in these lists for all $m\ge \KS(x)$. Consider the logarithm of the number of strings that follow $x$ in the $m$-th list.  We get a function that is defined for all $m\ge \KS(x)$ with $O(\log m)$ precision. The following result shows that this function describes the stochasticity profile of $x$ in different coordinates.

\begin{theorem}\label{th:position-deficiency}
Let $x$ be a string of length at most $n$.

\textup{(\textbf{a})}~Assume that $x$ appears in the list of strings of complexity at most $m$ and there are at least $2^k$ strings after $x$ in the list.  Then the pair $((m-k)+O(\log n),m-\KS(x))$ belongs to  the set $O(x)$.

\textup{(\textbf{b})}~Assume that the pair $(m-k,m-\KS(x))$ belongs to $O(x)$. Then $x$ appears in the list of strings of complexity at most $m+O(\log n)$ and there are at least $2^{k-O(\log n)}$ strings after it.
\end{theorem}

By Theorem~\ref{th:stochastic-optimal} the same statement holds for the set $S(x)$ in place of $O(x)$.

\begin{figure}
\begin{center}
\includegraphics[width=0.3\textwidth]{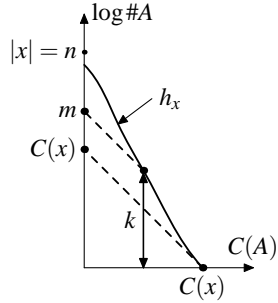}
\end{center}
\caption{To find how many strings appear after $x$ in the list of all strings of complexity at most $m$, we draw a line starting at $(0,m)$ with slope $-1$ and intersect it with the graph of $h_x$; if the second coordinate of the intersection point is $k$, there are about $2^{k}$ strings after $x$ in this list.}
\label{curve2}
\end{figure}

Ignoring the logarithmic correction and taking into account the relation between $O(x)$ and $h_x$, one can illustrate the statement of Theorem~\ref{th:position-deficiency} by Figure~\ref{curve2}.

\section{Time-bounded complexity and busy beavers}\label{depth}

There is one more way to get the stochasticity profile curve. Let us bound the computation time (number of steps) in the definition of Kolmogorov complexity and define $\KS^t(x)$ as the minimal length of a program that produces $x$ in at most $t$ steps. Evidently, $\KS^t(x)$ decreases as $t$ increases, and ultimately reaches $\KS(x)$.\footnote{One may ask which computational model is used to measure the computation time, and complain that the notion of time-bounded complexity may depend on the choice of an optimal programming language (decompressor) and its interpreter. Indeed this is the case, but we will use very rough measure of computation time based on busy beaver function, and the difference between computational models does not matter. The reader may assume that we fix some optimal programming language, and some interpreter (say, a Turing machine) for this language, and count the steps performed by this interpreter.} However, the convergence speed may be quite different for different $x$ of the same complexity. It is possible that for some $x$ the programs of minimal length produce $x$ rather fast, while other $x$ can be compressible only if we allow very long computations. Informally, the strings of the first type have some simple internal structure that allows us to encode them efficiently with a fast decoding algorithm, while the strings of the second type have ``deep'' internal structure that is visible only if the observer has a lot of computational power.

We use the so-called ``busy beaver numbers'' as landmarks for measuring the computation time. Let $BB(n)$ be the maximal running time of all programs of length at most $n$ (we use the programming language that defines Kolmogorov complexity, and some fixed interpreter for it).\footnote{Usually $n$-th busy beaver number is defined as the maximal running time or a maximal number of non-empty cells that can appear after Turing machine with at most $n$ states terminates starting on the empty tape. This gives a different number; we modify the definition so it does not depend on the peculiarities of encoding information by transition tables of Turing machines.} One can show that numbers $BB(n)$ have equivalent definition in terms of Kolmogorov complexity: \emph{$BB(n)$ is the maximal integer that has complexity at most $n$}. (More precisely, if $B(n)$ is the maximal integer that has complexity at most $n$, then $B(n-c)\le BB(n)\le B(n+c)$ for some $c$ and all $n$, and we ignore $O(1)$-changes in the argument of the busy beaver function.)

Now for every $x$ we may consider the decreasing function $i\mapsto \KS^{BB(i)}(x) - \KS(x)$ (it decreases fast for ``shallow'' $x$ and slowly for ``deep'' $x$; note that it becomes close to $0$ when $i=\KS(x)$, since then every program of length at most $\KS(x)$ terminate in $BB(\KS(x))$ steps.). The graph of this function is (with logarithmic precision) just a stochasticity profile, i.e., the set of points above the graph coincides with $O(x)$ 
up to a $O(\log n)$ error term:

\begin{theorem} \label{th-bauwens}
Let $x$ be a string of length  $n$.

\textup{(\textbf{a})}~
If a pair $(\alpha,\beta)$ is in $O(x)$, then $$\KS^{BB(\alpha+O(\log n))}(x)\le \KS(x)+\beta+O(\log n).$$

\textup{(\textbf{b})}~If  $\KS^{BB(\alpha)}(x)\le \KS(x)+\beta$, then the pair $(\alpha+O(\log n),\beta+O(\log n))$
is in $O(x)$.

\end{theorem}

By Theorem~\ref{th:stochastic-optimal} the same statement holds for the set $S(x)$ in place of $O(x)$. 


\section{How the stochasticity profile can look like?}

We have seen four different definitions that lead to the same (with logarithmic precision) notion of stochasticity profile. We see now that finite objects (strings) not only could have different complexities, but also the strings with the same complexity can be classified according to their stochasticity profiles.

However, we do not know yet that this classification is non-trivial: what if all strings of given complexity have the same stochasticity profile? The following result answers this question by showing that every simple decreasing function appears as complexity profile of some string.

\begin{theorem}\label{th:curves}
Assume that some integers $n$ and $k\le n$ are given, and $h$ is a non-increasing function mapping $\{0,1,\ldots,k\}$ 
to $\{0,1,\ldots,n-k\}$. Then there exists a string $x$ of length $n+O(\log n)+O(\KS(h))$ and complexity $k+O(\log n)+O(\KS(h))$ for which the set $O(x)$ 
\textup(and hence the set $S(x)$\textup)
coincides with the upper-graph of $h$ \textup(the set $\{\langle i,j\rangle\mid j\ge h(i) \text{ or } i\ge k\}$\textup) with $O(\log n+\KS(h))$
accuracy.
\end{theorem}

Note that the error term  depends on the complexity of $h$. If we consider simple functions $h$, this term is absorbed by our standard error term $O(\log n)$. In particular, this happens in two extreme cases: for the function $h\equiv 0$ and the function $h$ that is equal to $n-k$ everywhere. In the first case it is easy to find such a ``shallow'' $x$: just take an incompressible string of length $k$ and add $n-k$ trailing zeros to get a $n$-bit string. For the second case we do not know a better example than the one obtained from the proof of Theorem~\ref{th:curves}.

Let us say informally that a string $x$ of length $n$ is ``stochastic'' if its stochasticity profile $S(x)$ is close to the maximal possible set (achieved by the first example) with logarithmic precision, i.e., $x$ is $(O(\log n),O(\log n))$-stochastic. We know now that non-stochastic objects exist in the mathematical sense; a philosopher could ask whether they appear in the ``real life''. Is it possible that some experiment gives us data that do not have any adequate statistical model? This question is quite philosophical since having an object and a model we cannot say for sure whether the model is adequate in terms of algorithmic statistics. For example, the current belief is that the coin tossing data are described adequately by a fair coin model. Still it is possible that future scientists will discover some regularities in the very same data, thus making this model unsuitable.

We discuss the properties of stochastic objects in the next section. For now let us note only that this notion remains essentially the same if we consider probability distributions (and not finite sets) as models. Let us explain what does it mean.

Consider a probability distribution $P$ on a finite set of strings with rational values. It is a constructive object, so we can define the complexity of $P$ using some computable encoding. The conditional complexity $\KS(\cdot\cnd P)$ can be defined in the same way. Let us modify the definition of stochasticity and say that a string $x$ is ``$(\alpha,\beta)$-p-stochastic'' if there exists a distribution $P$ of the described type such that 
\begin{itemize}
\item $\KS(P)$ is at most $\alpha$;
\item $d(x\cnd P)$, defined as $-\log P(x) - \KS(x\cnd P)$, does not exceed $\beta$.
\end{itemize}
This is indeed a generalization: if $P$ is a uniform distribution, then the complexity of $P$ is (up to $O(1)$) the complexity of its support $A$, the value of $-\log P(x)$ is $\log\#A$, and using $P$ and $A$ as conditions gives the same complexity up to $O(1)$. On the other hand, this generalization leads only to a logarithmic change in the parameters:

\begin{theorem}\label{th:p-stochastic}
If some string $x$ of length $n$ is $(\alpha,\beta)$-p-stochastic, then the stting $x$ is also $(\alpha+O(\log n),\beta+O(\log n))$-stochastic.
\end{theorem}

Since all our statements are made with $O(\log n)$-precision, we may identify stochasticity with p-stochasticity (as we do in the sequel).

\section{Canonical models}

Let $\Omega_m$ denote the number of strings of complexity at most $m$. Consider its binary representation, i.e., the sum
    $$\Omega_m=2^{s_1}+2^{s_2}+\ldots+2^{s_t}, \text{ where $s_1>s_2>\ldots>s_t$.}$$
According to this decomposition, we may split the list itself into groups: first $2^{s_1}$ elements, next $2^{s_2}$ elements, etc.\footnote{We assume that an algorithm is fixed that, given $m$, enumerates all strings of complexity at most~$m$ in some order.} If $x$ is a string of complexity at most $m$, it belongs to some group, and this group can be considered as a model for $x$.

We may consider different values of $m$ (starting from $\KS(x)$). In this way we get different models of this type for the same $x$. Let us denote by $B_{m,s}$ the group of size $2^s$ that appears in the $m$-th list. Note that $B_{m,s}$ is defined only for $s$ that correspond to ones in the binary representation of $\Omega_m$. The models $B_{m,s}$ are called \emph{canonical} models in the sequel. The parameters of $B_{m,s}$ are easy to compute: the size is $2^s$ by definition, and the complexity is $m-s+O(\log m)$.

\begin{theorem}\label{th:canonical}
\textup{(\textbf{a})}~Every canonical model for a string $x$ lies on the boundary of $O(x)$ \textup(i.e., its parameters cannot be improved more than by $O(\log n)$ where $n$ is the length of $x$\textup).

\textup{(\textbf{b})}~For every point in $O(x)$ there exists a canonical model that has the same or better parameters \textup(with $O(\log n)$ precision\textup).
\end{theorem}
 
The second part of this theorem says that for every model $A$ for $x$ we can find a canonical model $B_{m,s}$ that has the same (or smaller) optimality deficiency, and $\KS(B_{m,s})\le \KS(A)$ with logarithmic precision. In fact, the second part of this statement can be strengthened: not only $\KS(B_{m,s})\le \KS(A)$, but also $\KS(B_{m,s}\cnd A)=O(\log n)$.

This result shows that (in a sense) we may restrict ourselves to canonical models. This raises the question: what are these models? What information they contain? The answer is a bit confusing: the information in models $B_{m,s}$ depends on $m-s$ only and is the same as the information in $\Omega_{m-s}$, the number of strings of complexity at most $m-s$:
\begin{theorem}\label{th:canonical-omega}
For all models $B_{m,s}$ both conditional complexities   $\KS(B_{m,s}\cnd \Omega_{m-s})$ and\\ $\KS(\Omega_{m-s}\cnd B_{m,s})$ are $O(\log m)$.
\end{theorem} 

One could note also that the information in $\Omega_k$ is a part of the information in $\Omega_l$ for $l\ge k$ (i.e., $\KS(\Omega_k\cnd \Omega_l)=O(\log l)$).\footnote{In fact, $\Omega_k$ contains the same information (up to $O(\log k)$ conditional complexity in both directions) as first $k$ bits of Chaitin's $\Omega$-number (a lower semicomputable random real), so we use the same letter $\Omega$ to denote it.}

Now it seems that finding a good model for $x$ does not provide any specific information about $x$: all we get (if we succeed) is the information about the number of terminating programs of bounded length, which that had nothing to do with $x$ and is the same for all $x$.

It is not clear how this philosophical collision between our goals and our achievements can be resolved. One of the approaches is to consider \emph{total conditional complexity}. This approach still leaves many questions open, but let us shortly describe it nevertheless.

We have said that ``string $a$ and $b$ contain essentially the same information'' if both $\KS(a\cnd b)$ and $\KS(b\cnd a)$ are small. This, however, does not guarantee that the properties of $a$ and $b$ are the same. For example, if $x^*$ is the shortest program for some highly non-stochastic string $x$, the string $x^*$ itself is perfectly stochastic.

To avoid this problem, we can consider total condition complexity $\KT(a\cnd b)$ defined as  the minimal length of a \emph{total} program $p$ such that $p(b)=a$. Here $p$ is called total if $p(b')$ halts for all $b'$, not only for $b$.\footnote{As usual, we assume that the programming language is optimal, i.e., gives $O(1)$-minimal value of the complexity compared to other languages.} This total conditional complexity can be much bigger than the standard conditional complexity $\KS(a\cnd b)$. It has the following property: if both $\KT(a\cnd b)$ and $\KT(b\cnd a)$ are negligible, there exists a computable permutation of low complexity that maps $b$ to $a$, and therefore the sets $O(a)$ and $O(b)$ are close to each other. (See~\cite{game-interpretation} for more details.)

Using this notion, we may consider a set $A$ as a ``strong'' model 
if it is close to the boundary of $O(x)$ and at the same time the \emph{total} complexity $\KT(A\cnd x)$ is small.  The second condition is far from trivial: one can prove that for some strings $x$ such strong models do not exist at all (except for the trivial model $\{x\}$ and the models of very small complexity)~\cite{ver13}.  But if strong models exists, they have some nice properties:  for example, the stochasticity profile of every strong sufficient  statistic for $x$ is close to the profile of the string $x$ itself~\cite{ver09}. (A model is called a sufficient statistic for $x$ if the optimality deficiency is small, i.e.,  the sum of its complexity and log-cardinality is close to $\KS(x)$.) The class of all sufficient statistics for  $x$ does not have this property (for some $x$).

Returning to the stochasticity profile, let us mention one more non-existence result. Imagine that we want to find a place when the set $O(x)$ touches the horizontal coordinate line. To formulate a specific task, consider for a given string of length $n$ two numbers. The first, $\alpha_1$, is the maximal value of $\alpha$ such that $(\alpha,0.1n)$ does not belong to $O(x)$; the second, $\alpha_2$, is the minimal value of $\alpha$ such that $(\alpha,10\log n)$ belongs to $O(x)$. (Of course, the constant $10$ is chosen just to avoid additional quantifiers, any sufficiently large constant would work.) Imagine that we want, given $x$ and $\KS(x)$, to find some point in the interval $(\alpha_1,\alpha_2)$, or even in a slightly bigger one (say, adding the margin of size $0.1n$ in both directions).   One can prove that \emph{there is no algorithm that fulfills this task}~\cite{vv02}.

\section{Stochastic objects}

The philosophical questions about non-stochastic objects in the ``real world'' motivate several mathematical questions. Where do they come from? can we obtain a non-stochastic object by applying some (simple) algorithmic transformation to a stochastic one? Can non-stochastic objects appear (with non-negligible probability) in a (simple) random process? What are the special properties of non-stochastic objects? 

Here are several results answering these questions.

\begin{theorem} \label{th:stochasticity-conservation}
Let $f$ be a computable total function. If string $x$ of length $n$ is $(\alpha,\beta)$-stochastic, then $f(x)$ is $(\alpha+\KS(f)+O(\log n),\beta+\KS(f)+O(\log n))$-stochastic.
\end{theorem}

Here $\KS(f)$ is the complexity of the program that computes $f$. 

An important example: let $f$ the projection function that maps every pair $\langle x,y\rangle$ (its encoding) to $x$. Then we have $\KS(f)=O(1)$, so we conclude that each component of an $(\alpha,\beta)$-stochastic pair is $(\alpha+O(\log n),\beta+O(\log n))$-stochastic.

A philosopher would interpret  Theorem~\ref{th:stochasticity-conservation} as follows: \emph{a non-stochastic object cannot appear in a simple total algorithmic process} (unless the input was already non-stochastic). Note that the condition of totality is crucial here: for every $x$, stochastic or not, we may consider its shortest program $p$. It is incompressible (and therefore stochastic), and $x$ is obtained from $p$ by a simple program (decompressor).

If a non-stochastic object cannot be obtained by a (simple total) algorithmic transformation from a stochastic one, can it be obtained (with non-negligible probability) in a (simple computable) random process? If $P$ is a simple distribution on a finite set of strings with rational values, then $P$ can be used as a statistical model, so only objects $x$ with high randomness deficiency $d(x\cnd P)$ can be non-stochastic, and the set of all $x$ that have $d(x\cnd P)$ greater than some $d$ has negligible $P$-probability (an almost direct consequence of the deficiency definition). 

So for computable probabilistic distributions the answer is negative for trivial reasons. In fact, much stronger (and surprising) statement is true. Consider a probabilistic machine $M$ without input that, being started, produces some string and terminates, or does not terminate at all (and produces nothing). Such a machine determines a \emph{semimeasure} on the set of strings (we do not call it measure since the sum of probabilities of all strings may be less than $1$ if the machine hangs with positive probability). The following theorem says that a (simple) machine of this type produces non-stochastic objects with negligible probability.

\begin{theorem}\label{th:random-stochastic}
There exists some constant $c$ such that the probability of the event \begin{center}
``$M$ produces a string of length at most $n$ that is not $(d+\KS(M)+c\log n, c\log n)$-stochastic''
\end{center}
 is bounded by $2^{-d}$ for every machine $M$ of described type and for arbitrary integers $n$ and~$d$.
\end{theorem}

The following results partially explain why this happens. Recall that algorithmic information theory defines \emph{mutual information} in two strings $x$ and $y$ as $\KS(x)+\KS(y)-\KS(x,y)$; with $O(\log n)$ precision (for strings of length at most $n$) this expression coincides with $\KS(x)-\KS(x\cnd y)$ and $\KS(y)-\KS(y\cnd x)$. Recall that by $\Omega_n$ we denote the number of strings of complexity at most $n$. 

\begin{theorem}\label{th:nonstochasticity-information}
There exists a constant $c$ such that for every $n$, for every string $x$ of length at most $n$ and for every threshold $d$ the following holds: if a string $x$ of length $n$ is not $(d+c \log n, c\log n)$-stochastic, then 
        $$I(x:\Omega_n)\ge d-c\log n.$$
\end{theorem}

This theorem says that all non-stochastic objects have a lot of information about a specific object, the string $\Omega_n$. This explain why they have small probability to appear in a (simple) randomized process, as the following result shows. It guarantees that for every fixed string $w$ the probability to get (in a simple random process) some object that contains significant information about $w$, is negligible.

\begin{theorem}\label{th:no-information-appears}
There exists a constant $c$ such that for every $n$, for every probabilistic machine $M$, for every string $w$ of length at most $n$  and for every threshold $d$ the probability of the event 
\begin{center}
``$M$ outputs a string $x$ of length at most $n$ such that $I(x:w)>\KS(M)+d+c\log n$''
\end{center}
 is at most $2^{-d}$.
\end{theorem}

The last result of this section shows that stochastic objects are ``representative'' if we are interested only in the complexity of strings and their combinations: for every tuple of strings one can find a stochastic tuple that is indistinguishable from the first one by complexities of its components.

\begin{theorem}\label{th:typization}
For every $k$ there exists a constant $c$ such that for every $n$ and for every $k$-tuple $\langle x_1,\ldots,x_k\rangle$ of strings of length at most $n$, there exist another $k$-tuple $\langle y_1,\ldots, y_k\rangle$ that is $(c\log n, c\log n)$-stochastic and for every $I\subset\{1,2,\ldots,n\}$ the difference between $\KS(x_I$) and $\KS(y_I)$ is at most $c\log n$. 
\end{theorem}

Here $x_I$ is a tuple made of strings $x_i$ with $i\in I$; the same for $y_I$.

This result implies, for example, that every linear inequality for complexities that is true for stochastic tuples, is true for arbitrary ones.

However, there are some results that are known for stochastic tuples but still are not proven for arbitrary ones. See~\cite{muchnik-romashchenko} for details. 

\section{Restricted classes: Hamming balls as descriptions}

Up to now we considered arbitrary sets as statistical models. However, sometimes we have some external information that suggests a specific class of models (and it remains to choose the parameters that define some model in this class). For example, if the data string is a message sent through a noisy channel that can change some bits,  we consider Hamming balls as models, and the parameters are the center of this ball (the original message) and its radius (the maximal number of changed bits).

So let us consider some family $\mathcal{B}$ of finite sets. To get a reasonably theory, we need to assume some properties of this family:

\begin{itemize}

\item The family $\mathcal{B}$ is computably enumerable: there exists an algorithm that enumerates all elements of $\mathcal{B}$ (finite sets are here considered as finite objects,  encoded as lists of their elements).
 
\item For each $n$ the set of all $n$-bit strings belongs to $\mathcal{B}$.

\item There exists a polynomial $p$ such that the following property holds: for every $B\in\mathcal{B}$, for every positive integer $n$ and for every $c<\#B$ the set of all $n$-bit strings in $B$ can be covered by $p(n)\#B/c$ sets from $\mathcal{B}$ and each of the covering sets has cardinality at most $c$.

\end{itemize}

Here $\#B$ stands for the cardinality of $B$. Counting argument shows that in the last condition we need at least $\#B/c$ covering sets;  the condition says that polynomial overhead is enough here.

One can show (using simple probabilistic arguments) that the family of all Hamming balls (for all string lengths, centers and radii) has all three properties. This family is a main motivating example for our considerations.
\medskip

Now we can define the notion of a $\mathcal{B}$-$(\alpha,\beta)$-stochastic object: a string $x$ is $\mathcal{B}$-$(\alpha,\beta)$-stochastic  if there exists a set $B\in \mathcal{B}$ containing $x$ such that $\KS(B)\le\alpha$ and $d(x\cnd B)\le \beta$.  (The original notion of $(\alpha,\beta)$-stochasticity corresponds to the case when $\mathcal{B}$ contains all finite sets.)   For every $x$ we get a set $S_\mathcal{B}(x)$ of pairs $(\alpha,\beta)$ for which $x$ is $\mathcal{B}$-$(\alpha,\beta)$-stochastic.  We can also define the set $O_\mathcal{B}(x)$ using optimality deficiency instead of randomness deficiency. The $\mathcal{B}$-version of Theorem~\ref{th:stochastic-optimal} is still true (though the proof needs a much more ingenious construction):

\begin{theorem}\label{th:B-stochastic-optimal}
Let $\mathcal{B}$ be the family of finite sets that has the properties listed above. Then the for every string $x$ of length at most $n$ the sets $S_\mathcal{B}(x)$ and $O_\mathcal{B}(x)$ coincide up to a $O(\log n)$ error term.
\end{theorem}

The proof is more difficult (compared to the proof of Theorem~\ref{th:stochastic-optimal}) since we now need to consider sets in $\mathcal{B}$ instead of arbitrary finite sets. So we cannot construct the required model for a given string~$x$ ourselves and have to select it among the given sets that cover $x$. This can be done by a game-theoretic argument.

It is interesting to note that a similar argument can be used to obtain the following result about stochastic finite set (Epstein--Levin theorem):

\begin{theorem}
\label{th:epstein-levin}
If a finite set $X$ is $(\alpha,\beta)$-stochastic and the total probability 
$$\sum_{x\in X} 2^{-\KP(x)}$$
of its elements exceeds $2^{-k}$, then $X$ contains some element $x$ such that 
$$\KP(x)\le k+\KP(k)+\log\KP(k)+\alpha+O(\log\beta)+O(1).$$
\end{theorem}

Here $\KP(u)$ stands for the prefix complexity of $u$ (see, e.g.,~\cite{li-vitanyi} for the definition).  To understand the meaning of this theorem, let us recall one of the fundamental results of the algorithmic information theory: the (prefix) complexity of a string $x$ equals the binary logarithm of its a priori probability. If we consider a set $X$ of strings instead of one string $x$, we can consider the a priori probability of $X$ (expressing how difficult is to get some element of $x$ in a random process) and the minimal complexity of elements of $X$ (saying how difficult is to specify an individual element in $X$). The fundamental result mentioned above says that for singletons these two measures are closely related; for arbitrary finite sets it is no more the case, but Theorem~\ref{th:epstein-levin} guarantees that for the case for \emph{stochastic} finite sets.

\medskip
Returning to our main topic,  let us note that for Hamming balls the boundary curve of $O_\mathcal{B}(x)$ has a natural interpretation. To cover $x$ of length $n$ with a ball $B$ with center $y$ having cardinality $2^\beta$ and complexity at most $\alpha$ means (with logarithmic precision) to find a string $y$ of complexity at most $\alpha$ in the $r$-neighborhood of $x$, where $r$ is chosen is such a way that balls of radius $r$ have about $2^\beta$ elements.  So this boundary curve represents a trade-off between the complexity of $y$ and its distance to $x$.

Again one can ask what kind of boundary curves may appear. As in Theorem~\ref{th:curves}, we can get essentially arbitrary non-increasing function. However, here precision is worse: $O(\log n)$ term is now replaced by $O(\sqrt{n\log n})$.

\begin{theorem}\label{th:b-curves}
Assume that some integers $n$ and $k\le n$ are given, and $h$ is a non-increasing function mapping $\{0,1,\ldots,k\}$ to $\{0,1,\ldots,n-k\}$. Then there exists a string $x$ of length $n+O(\sqrt{n\log n})+O(\KS(h))$ and complexity $k+O(\sqrt{n\log n})+O(\KS(h))$ for which the set $O_\mathcal{B}(x)$ coincides with the 
upper-graph of $h$ \textup(the set $\{\langle i,j\rangle\mid j\ge h(i) \text{ or } i\ge k\}$\textup) with $O(\sqrt{n\log n}+\KS(h))$-precision.
\end{theorem}

Unlike the general case where non-stochastic objects (for which the curve is far from zero) exists but are difficult to describe, for the case of Hamming balls one can give more explicit examples. Consider some explicit error correction code that has distance $d$. Then every string that differs in at most $d/2$ positions from some codeword $x$, has almost the same complexity as $x$ (since $x$ can be reconstructed from it by error correction). So the balls of radius less than $d/2$ containing some codeword have almost the same complexity as the codeword itself (and the balls of zero radius containing it).

Let $x$ be a typical codeword of this binary code (its complexity is close to the logarithm of the number of codewords). For values of $\alpha$ slightly less than $\KS(x)$ we need a large $\beta$ (at least the logarithm of the cardinality of a ball of radius $d/2$) to make such a codeword $(\alpha,\beta)$-stochastic.

\section{Historical remarks}

The notion of $(\alpha,\beta)$-stochasticity was mentioned by Kolmogorov in his talks at the seminar he initiated at the Moscow State University in early 1980s (see~\cite{alpha-beta}). The equivalence between this notion and the optimality deficiency (Theorem~\ref{th:stochastic-optimal}) was discovered in~\cite{vv02}.

The connections between the existence of adequate models and the position in the list of strings of bounded complexity was discovered by G\'acs, Tromp and Vit\'anyi in~\cite{GTV}, though this paper considered only the position of $x$ in the list of strings of complexity at most $\KS(x)$. Theorems~\ref{th:position-invariance} and~\ref{th:position-deficiency} appeared in~\cite{vv02}. The paper~\cite{GTV} considered also the canonical models (called ``nearly sufficient statistics'' in this paper) for the case $m=\KS(x)$.  In the general case the canonical models were considered in~\cite{vv02} (section V, \emph{Realizing the structure function}), where Theorems~\ref{th:canonical} and~\ref{th:canonical-omega} were proven.

The minimal description length principle goes back to Rissanen~\cite{rissanen}; as he wrote in this paper, ``If we work with a fixed family of models, $\langle\ldots\rangle$ the cost of the complexity of a model may be taken as the number of bits it takes to describe its parameters.  Clearly now, when adding new parameters to the model, we must balance their own cost against the reduction they permit in the ideal code length, $-\log P(x\cnd\theta)$, and we get the desired effect in a most natural manner. If we denote the total number of bits required to encode the parameters $\theta$ by $L(\theta)$, the we can write the total code length as $L(x,\theta)=-\log P(x\cnd\theta)+L(\theta)$, which we seek to minimize over $\theta$''.  The set denoted by $O(x)$ in our survey was considered in 1974 by Kolmogorov (see~\cite{Ko74a}); later it appeared in the literature also under the names of ``sophistication'' and ``snooping curves''.

The notion of sophistication was introduced by Koppel in~\cite{koppel}. Let $\beta$ be a natural number; \emph{$\beta$-sophistication} of a string $x$ is the minimal length of a total program $p$ such that there is a string $y$ with $p(y)=x$ and $|p|+|y|\le \KS (x)+\beta$. In out terms $p$ defines a model that consists of all $p(y)$ for all strings $y$ of a given length. It is not hard to see that with logarithm precision we get the same notion: the $\beta$-sophistication of $x$ is at most $\alpha$ if and only if the pair $(\alpha,\beta)$ is in the set $O(x)$.

The notion of snooping curve $L_x(\alpha)$ of $x$ was introduced by V'yugin in~\cite{Vy01}. In this paper he considered strategies that read a bit sequence from left to right and for each next bit provide a prediction (a rational-valued probability distribution on the set $\{0,1\}$ of possible outcomes). After the next bit appears, the \emph{loss} is computed depending on the prediction and actual outcome. The goal of the predictor is to minimize the total loss, i.e., the sum of losses at all $n$ stages (for a $n$-bit sequence). Vyugin considered different loss functions, and for one of them, called \emph{logarithmic loss function}, we get a notion equivalent to $O(x)$. For a logarithmic loss function, we account for loss $-\log p$ if the predicted probability of the actual outcome was $p$. It is easy to see that for a given $x$ the following statement is true (with logarithmic precision): there exists a strategy of complexity at most $\alpha$ with loss at most $l$ if and only if $l\ge h_x(\alpha)$. (Indeed, prediction strategies are just bit-by-bit representation of probability distributions on the set of $n$-bit strings, in terms of conditional probabilities.)

Theorem~\ref{th-bauwens} (Section~\ref{depth}) is due to Bauwens~\cite{bauwens}. The idea to consider the difference between time bounded complexity of $x$ and the unbounded one goes back to Chaitin~\cite{chaitin77}. Later the subject was studied by Bennett who introduced the notion of logical depth: the  \emph{depth of $x$ at significance level $\beta$} is the minimal time $t$ such that $\KS^t(x)\le \KS(x)+\beta$. The string is called $(\beta,t)$-deep if its depth at significance level $\beta$ is larger than $t$. A closely related notion of computational depth was introduced in~\cite{antunes-et-al}: the \emph {computational  depth of $x$ with time bound $t$} is $\KS^t(x)-\KS(x)$. Obviously, computational  depth of $x$ with time bound $t$ is more than $\beta$ if and only if $x$ is $(\beta,t)$-deep. Theorem~\ref{th-bauwens} relates both notions of depth  to the stochasticity profile (with logarithmic precision): a string is $(\beta,B(\alpha))$-deep if and only if the pair $(\alpha,\beta)$ is outside the set~$O(x)$.

Theorem~\ref{th:curves} was proved in~\cite{vv02}. Long before this paper (in 1987)  V'yugin established that set $S(x)$ can assume all possible shapes (within the obvious constraints) but only for $\alpha = o(|x|)$. Also, according to Levin~\cite{Le02}: ``Kolmogorov told me about $h_x (\alpha)$  and asked how it could behave. I proved that $h_x (\alpha) + \alpha +O(\log \alpha)$ is monotone but otherwise arbitrary within $\pm O(p \log \alpha)$ accuracy where $p$ is the number of ``jumps'' of the arbitrary function imitated; it stabilizes on $\KS(x)$ when $\alpha$ exceeds $I(\chi : x)$ [the information in the characteristic sequence $\chi$ of the ``halting problem'' about $x$]. The expression for accuracy was reworded by Kolmogorov to $O(\sqrt{\alpha \log \alpha})$ [{\em square root accuracy}];  I gave it in the above, less elegant, but equivalent, terms. He gave a talk about these results at a meeting of Moscow Mathematical Society \cite{Ko74b}.'' This claim of Levin implies  Theorem~\ref{th:nonstochasticity-information} that was published in~\cite{vv02}. 
 
Theorem~\ref{th:p-stochastic} (mentioned in~\cite{alpha-beta}) is easy and Theorem~\ref{th:stochasticity-conservation} easily follows from Theorem~\ref{th:curves}.

The existence of non-$(\alpha,\beta)$-stochastic strings (for small $\alpha, \beta$)
was mentioned in~\cite{alpha-beta}. Then V'yugin~\cite{Vy85} and Muchnik~\cite{muchnik-meta} showed that that their a priori measure is about $2^{-\alpha}$, a direct corollary of which is our Theorem~\ref{th:random-stochastic}.

Theorems~\ref{th:nonstochasticity-information} and \ref{th:no-information-appears} are essentially due to Levin (see~\cite{Le02} and~\cite{levin84}). 

Theorem~\ref{th:typization} is easy to prove using A.~Romashchenko's ``typization'' trick (see~\cite{inequalities,combinatorial-interpretation}).

Theorems~\ref{th:B-stochastic-optimal} and \ref{th:b-curves} appeared in~\cite{vv10};  Theorem~\ref{th:epstein-levin} appeared in~\cite{epstein-levin}.

\textbf{Acknowledgements}.
Authors thank all their colleagues with whom they discussed algorithmic statistics, especially Bruno~Bauwens, Laurent Bienvenu, Sam~Epstein, Peter G\'acs,  Leonid Levin, Paul Vit\'anyi, Vladimir V'yugin, and all the members of the Kolmogorov seminar (Moscow) and ESCAPE team (Marseille, Montpellier).

The preparation of this survey was supported in part by the EMC ANR-09-BLAN-0164 and RFBR 12-01-00864 grant.

\end{document}